\documentclass[graybox]{svmult}

\usepackage{type1cm}        
\usepackage{makeidx}         
\usepackage{graphicx}        
\usepackage{multicol}        
\usepackage[bottom]{footmisc}

\usepackage{newtxtext}       %
\usepackage[varvw]{newtxmath}       

\makeindex             

\begin{document}

\title*{Simulations of two-dimensional 
single-mode Rayleigh--Taylor Instability using front-tracking/ghost-fluid method: comparison to 
experiments and theory}
\titlerunning{FT/GFM method for RTI simulations}
\author{James Burton \and Tulin Kaman\orcidID{0000-0003-2743-0951}}
\institute{
James Burton \at Department of Mathematical Sciences, 
University of Arkansas, Fayetteville, AR 72701, USA. \email{jmb100@uark.edu}
\and Tulin Kaman \at 
Department of Informatics, University of Zurich, Zurich, CH-8050, Switzerland.\\
Department of Mathematical Sciences, 
University of Arkansas, Fayetteville, AR 72701, USA.
\email{tulin.kaman@gmail.com}}
%
%
\maketitle
\abstract{
Two-dimensional single-mode Rayleigh--Taylor Instability (RTI)
is simulated using an accurate and robust front-tracking/ghost-fluid method (FT/GFM) with high-order weighted essentially non-oscillatory (WENO) scheme.
We compare our numerical results with the single-mode RTI experiments of Renoult~{\it et al.}~\cite{RenRosCar15}.
The time evolution of the interface between two immiscible fluids 
and the effects of surface tension on the growth of the amplitude and asymmetry of the perturbed interface are examined for the initial wavelength $\lambda=1$~cm and the Atwood number $A=0.29$.
The important features of RTI flows such as interface profiles, bubble/spike penetration and velocities show good agreement between experiments and simulations of immiscible fluids with surface tension. 
The velocity vector fields for the bubble and spike in the linear and nonlinear regimes are consistent with the theory for the single wavelength perturbation.}

\keywords{Rayleigh--Taylor Instability; Front Tracking; 
Ghost-Fluid Method}

\section{Introduction}
\label{sec:intro}
The Rayleigh--Taylor instability (RTI) is a classical hydrodynamical instability driven by an acceleration force applied across a density discontinuity.
The light fluid penetrates the heavy fluid in bubbles, 
and the heavy fluid penetrates the light fluid in spikes.
The penetration distance of the light fluid into the heavy fluid, known as bubble penetration, is described by $ h_{b} = \alpha A g t^2 $
where $\alpha$ is the growth parameter, the Atwood number $A = (\rho_1 - \rho_2)/(\rho_1 + \rho_2)$ is the dimensionless measure of density contrast, $g$ is the acceleration and $t$ is the time. 
Many theoretical, experimental and computational studies 
have been performed to understand the dynamics 
of turbulent mixing due to the RTI and 
important features such as interface dynamics and bubble/spike growth parameters have been investigated.
Abarzhi~\cite{Aba10} reviews theoretical approaches and highlights key features of turbulent mixing due to RTI. 
Banerjee~\cite{Ban20} presents a comprehensive overview of experimental designs and important diagnostics. 
Additionally, the book on hydrodynamic instabilities and turbulence~\cite{Zho24}, and review articles by Zhou \cite{Zho17a,Zho17b}, provide detailed discussions of experimental, theoretical, and computational studies of turbulent mixing arising from Rayleigh--Taylor, Richtmyer--Meshkov, and Kelvin--Helmholtz instabilities.


The earlier computational studies of RTI simulations by Glimm~{\it et al.}~\cite{KamGliSha10, KamMelRao11, GliShaKam11, ZhaKamShe18, GliCheSha20} 
agreed with the experiments of Read~\cite{Rea84}, Smeeton \& Youngs~\cite{SmeYou87}, Ramaprabhu \& Andrews~\cite{RamAnd04} and Muesche~\cite{Mue08}.
In these RTI simulations, the filtered Navier-Stokes equations for immiscible and miscible fluids in an inertial frame were solved
using front tracking and large-eddy simulations  with subgrid-scale methods.  
Recently, an increasingly accurate and robust front-tracking (FT) method coupled with the ghost-fluid method (GFM) and high-order weighted essentially non-oscillatory (WENO) schemes 
was proposed for the numerical simulations of single-mode Richtmyer-Meshkov Instability (RMI) of an air/SF$_6$ interface~\cite{HolKam25, KamHol22}. 
RMI simulations based on FT/GFM with high-order WENO showed improvements 
in the late-time dynamics of the interfaces.
We presented a good agreement with the single-mode shock tube experiments of Collins and Jacobs 2002~\cite{ColJac02} for incident shock strength $M=1.21$~\cite{HolKam25}. 
These methods are implemented in the \textsc{FronTier} software package, which is extended to handle surface tension effects for RTI simulations of immiscible fluids presented here. 

In this study, we investigate the time evolution of two-dimensional single-mode RTI between 
two immiscible fluids using FT/GFM with the classical fifth-order WENO. 
We performed new single-mode immiscible RTI simulations to investigate the effects of surface tension
on the growth of the amplitude and asymmetry of the perturbed interface in the linear and nonlinear regimes. 
The experiments of Renoult~{\it et al.}~\cite{RenRosCar15} are used as benchmark tests for the validation of our numerical simulations. 
Renoult~{\it et al.}~\cite{RenRosCar15} conducted RTI experiments using a magnetic levitation technique to impose a precise single-mode sinusoidal initial disturbance at the interface between two immiscible fluids.
\textsc{FronTier} is used to capture the exact location of the interface between fluids. 
We compare our numerical results with the experiments of Renoult~{\it et al.}~\cite{RenRosCar15} for the Atwood number $A=0.29$, where two fluids are subject to Earth's gravity $g_0$. 
In addition, we investigate the velocity vector fields for the spike and bubble and compare them with the theory of Abarzhi~\cite{ChaJaiHwa23}.
The organization of this paper is as follows. 
In Sect.~\ref{sec:alg}, the algorithmic framework used in this study is briefly described. 
In Sect.~\ref{sec:exp}, the Renoult~{\it et al.}~\cite{RenRosCar15} experimental setup is introduced. 
In Sect.~\ref{sec:resuRTI}, we present our numerical results and compare simulation data with experimental data and theory. 

\section{Algorithmic Framework}
\label{sec:alg}

In this section, front-tracking (FT) coupled with the ghost-fluid method (GFM) for the discretization of the governing equations, the two-dimensional 
compressible Euler equations~(\ref{eq:Euler}) for inviscid fluids, is described.

\begin{equation}
\label{eq:Euler}
{\mathbf U}_t + {\mathbf F} ( {\mathbf U})_x + {\mathbf G} ( {\mathbf U})_y={\mathbf 0}
\end{equation}
with 
\begin{equation}
{\bf U}=\begin{pmatrix}
{\rho} \\
{\rho} {u} \\ 
{\rho} {v} \\ 
{E} \\
\end{pmatrix}, 
{\bf F} ( {\bf U})=
\begin{pmatrix}
{\rho} {u}\\
{\rho} {u}^2 + {p} \\
{\rho} u v \\
({E}+{p})u \\
\end{pmatrix}, 
{\bf G} ( {\bf U})=
\begin{pmatrix}
{\rho} v\\
{\rho} u v \\
{\rho} v^2 + p \\
({E}+{p})v \\
\end{pmatrix} \nonumber
\end{equation} 
where ${\bf U}$ is the vector of conserved variables (mass, momentum, energy)
and ${\bf F}( {\bf U})$ and ${\bf G} ( {\bf U})$ are the fluxes.
Here $\rho$ is the density, $(u, v)$ is the velocity in $(x,y)$ directions, $p$ is the pressure, 
$E=\rho e + \frac{1}{2} \rho (u^2+v^2)$ is the total energy,
$e=\frac{p} {(\gamma - 1) \rho }$ is the specific internal energy and
$\gamma$ is the constant specific heat ratio.

FT enforces the exact fluid jump conditions at tracked interfaces by solving the Riemann problem.
The interfaces that are boundaries between the two fluids are discretized by assigning special degrees of freedom to surfaces or lower-dimensional manifolds.
The unwanted mixing is prevented by explicitly tracking the interfaces as they move dynamically through a background rectangular grid.
GFM avoids the unphysical oscillations at the fluid interfaces by separating the domain for each fluid and defining each fluid domain with a ghost-fluid region. 
Physical quantities across the interface are defined in the ghost region to ensure continuity and appropriate boundary conditions.
The governing equations are solved independently in each fluid domain. 
Although Fedkiw's original GFM~\cite{FedOsh99} is a robust method, spurious pressure oscillations and large diffusion errors are reported for multiphase flows with high density and pressure ratios.
The unphysical pressure oscillations for compressible gas-water simulations were overcome by Liu~\emph{et al.}~\cite{LiuKhoWan05}, who developed the modified GFM by using a Riemann solver to construct the ghost fluid state in the computational domain.
The idea of combining FT with GFM was first proposed by Terashime and Tryggvason~\cite{TerTry09}.
Later, FT of Glimm~\emph{et al.}~\cite{GliGroLi99a} was combined with GFM and used for numerical simulations of the primary breakup of a liquid jet~\cite{BoLiuGli10}.

We extended FT/GFM with high-order WENO for numerical simulations of shock-induced turbulent mixing~\cite{HolKam25}. 
The high-order WENO are numerical methods for solving hyperbolic conservation laws, especially for problems with shocks and discontinuities~\cite{Shu20}.
The choice of stencils and their weights play an important role in the stability and accuracy of the WENO schemes. 
A class of WENO schemes with increasingly high-order accuracy are designed to ensure that unphysical spurious oscillations are minimized near discontinuities and maximal order of accuracy is achieved in smooth regions~\cite{BalShu00}.

\section{Experiments}
\label{sec:exp}
The single-mode RTI experiments of Renoult~{\it et al.}~\cite{RenRosCar15} are used for validation and verification studies to quantify the reliability of our new code.  
Renoult~{\it et al.}~\cite{RenRosCar15} 
performed RTI experiments using a
magnetic levitation technique to create a single-mode 
sinusoidal initial perturbation of the two-dimensional interface between two immiscible fluids.
The heavy fluid $\rho_1= 1.398~\text{g}/\text{cm}^{3}$ is a paramagnetic aqueous mixture, and the light fluid $\rho_2=0.773~\text{g}/\text{cm}^{3}$ is hexadecane.
The low Atwood number $A = (\rho_1 - \rho_2)/(\rho_1 + \rho_2) = 0.29$ makes it less costly to simulate with our compressible 
code. 
The interface between these two immiscible fluids of different densities is subject to gravity $g_0$ and interfacial tension $\gamma=7.74~\text{dyn}\cdot\text{cm}^{-1}$. 
In their experiments, they considered a small initial perturbation, 
where the ratio between the initial amplitude $a_0$ and the initial wavelength $\lambda_0$ is $a_0/\lambda_0 \ll 1$. The experimental runs were performed for five wavelengths $\lambda_0 = 1, 1.25, 1.5 , 1.75$ and $2$~cm. 
Two dimensionless metrics $M_{AE}=(a^{LR} - a)/\lambda $ and $M_{SE}=\Delta x_b/\Delta x_s -1 $ 
were defined to measure the amplitude effects and the asymmetry effects of nonlinearities.
Here, $a^{LR}$ is the predicted amplitude based on a linear regime, $a$ is one-half the measured peak-to-peak amplitude, $\lambda$ is the fundamental wavelength and $\Delta x_b$ and $\Delta x_s$ are the distance between two nodes that separate a rising bubble and a falling spike.

\section{Numerical Results}
\label{sec:resuRTI}

In our numerical studies, the computational domain is set to 
$[0,L_x] \times [-L_y,L_y]$ where $L_x=3$~cm and $L_y=4$~cm.
The perturbed interface at $y=0$~cm
is initialized on a $3\times 8$ computational domain that 
has reflection symmetry boundary conditions in the horizontal direction and no-flux boundary conditions in the vertical direction. 
The initial sinusoidal perturbed interface amplitude $a_0$ is chosen 
such that $ka_0 = 0.27$ where $k$ is the wave number of the perturbation. 
The initial amplitude of the perturbation is $9\%$ of its wavelength. 
The surface tension between two immiscible fluids in the RTI simulations 
is $\sigma= 7.74~\text{dyn}\cdot\text{cm}^{-1}$
according to the experiments of Renoult~{\it et al.}~\cite{RenRosCar15}. 
We focus on the particular initial perturbation wavelength  $\lambda_0=1$~cm to compare the interface profiles in the experimental images. 
The grid resolution $192\times 512$ is chosen so that there are $64$ mesh points per initial perturbation wavelength.

\begin{figure}
\centering
\includegraphics[scale=.3]{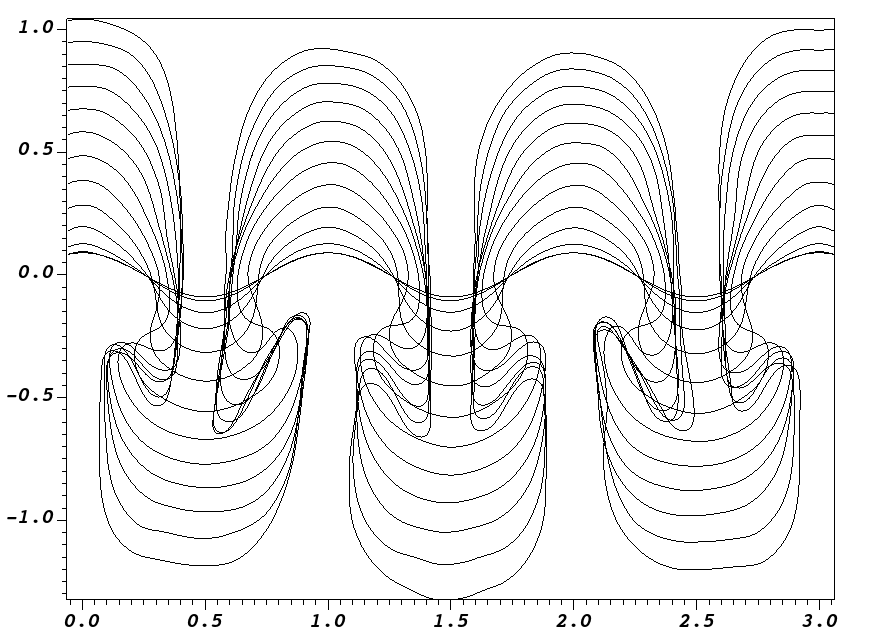}
\caption{Interface profile showing the development of immiscible RTI with a surface tension $\sigma= 7.74~\text{dyn}\cdot\text{cm}^{-1}$ and an initial wavelength $\lambda_0=1$~cm at a grid resolution $192 \times 512$.
There is a 16.7~ms between two consecutive interface.}
\label{fig:intfc-k3}
\end{figure}

\begin{figure}
\centering
\includegraphics[scale=.4]{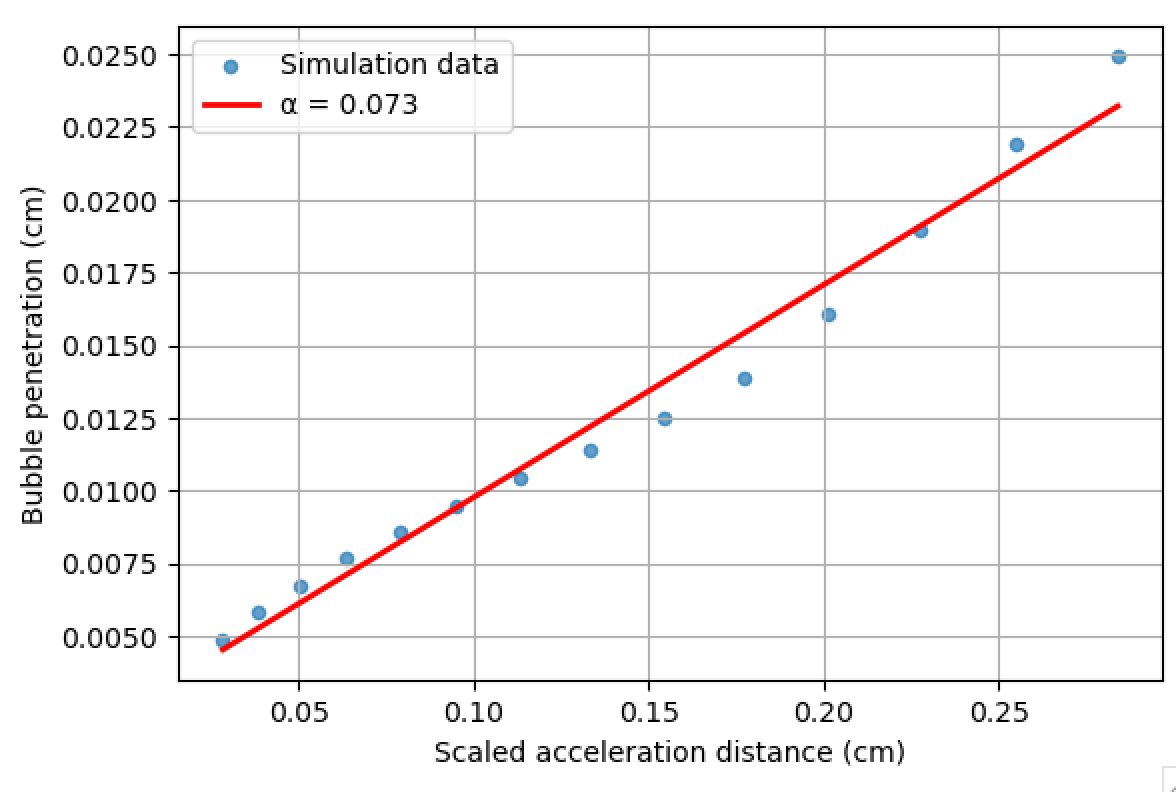}
\caption{
Bubble penetration versus scaled acceleration distance for $\lambda_0=1$~cm at a grid resolution $192\times 512$. The growth parameter $\alpha=0.073$ is obtained by fitting a line to the numerical simulation data.}
\label{fig:hb-X} 
\end{figure}

Fig.~\ref{fig:intfc-k3} shows the interface front between the heavy (top) and light (bottom) fluids 
for the Atwood number $A=0.29$.
The interface plots were generated every $16.7$~ms.
While the distance between two nodes $\Delta x_b$ of the rising bubbles increases, the distance between two nodes 
$\Delta x_s$ of the falling spike decreases as time proceeds. 
In the linear regime at early times ($t<0.6$), the bubble and spikes are symmetric. 
The asymmetry on the bubble and spike structures of the instability starts to develop 
as RTI transitions from the linear to the nonlinear regime.
This transition affects the exponential growth of the amplitude and symmetry of the interface.
Fig~\ref{fig:hb-X} shows the bubble penetration $h_b$ versus the scaled acceleration distance $X=Agt^2$ of 
the numerical simulation data. 
The growth parameter $\alpha=0.073$ is obtained by fitting a line to the numerical simulation data.
The RTI growth parameter computed for the numerical simulation of Renoult~{\it et al.}~\cite{RenRosCar15} experiment for wavelength $\lambda =1$~cm agrees well with previously reported results for other immiscible experiments~\cite{GliShaKam11}. 

\begin{figure}
\centering
\includegraphics[scale=.35]{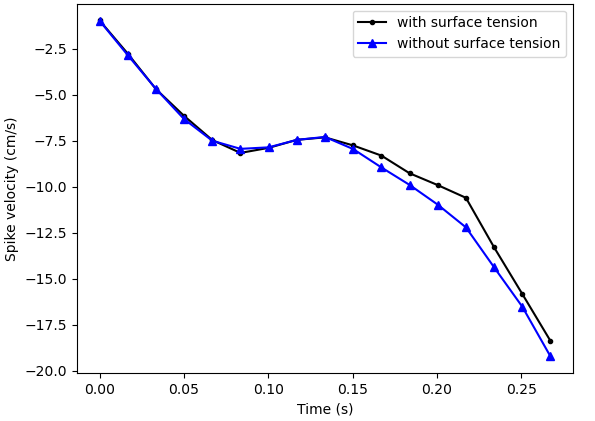}
\includegraphics[scale=.35]{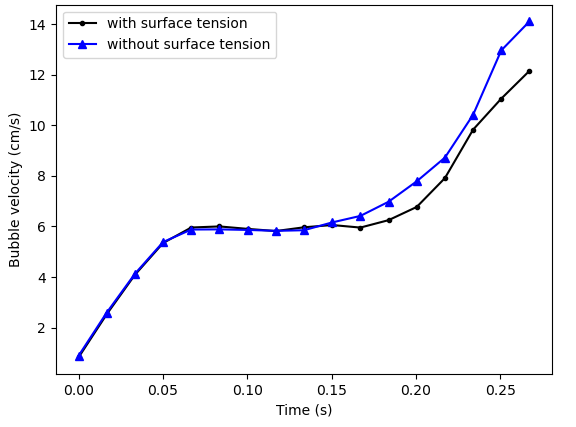}\\
\includegraphics[scale=.35]{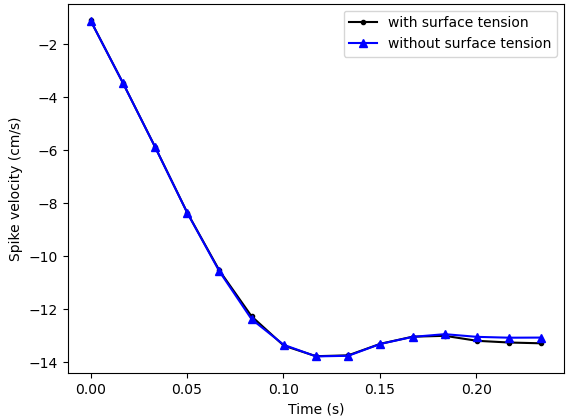}
\includegraphics[scale=.35]{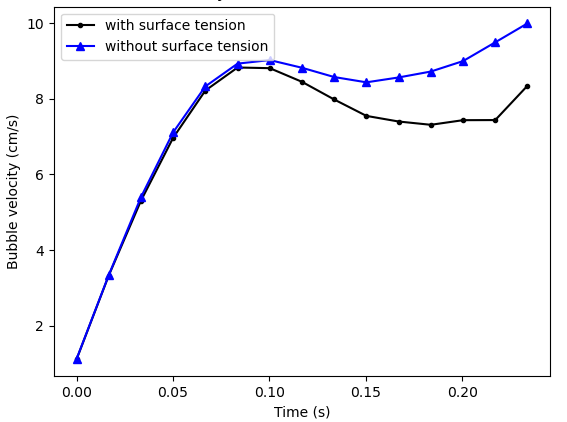} 
\caption{Comparison of spike (left) and bubble (right) velocities for solutions of the compressible Euler equations with and without a surface tension on a fine grid resolution $192 \times 512$ for 
3 waves (top) and single wave (bottom).}
\label{fig:velocities}
\end{figure}

Fig.~\ref{fig:velocities} shows the spike and bubble velocities measured at the tip of the spike (left) and the tip of the bubble (right) with and without interfacial tension. The effect of interfacial tension is not observed on the spike and bubble velocities for the initial wavelength $\lambda=1$~cm (top) and 
$\lambda=3$~cm (bottom) in the linear regime. 
However, we clearly see the effect of surface tension on the bubble velocity for a single wavelength perturbed interface in the transitional regime from linear to nonlinear (see bottom right in Fig.~\ref{fig:velocities}). 

\begin{figure}
\centering
\includegraphics[scale=.42]{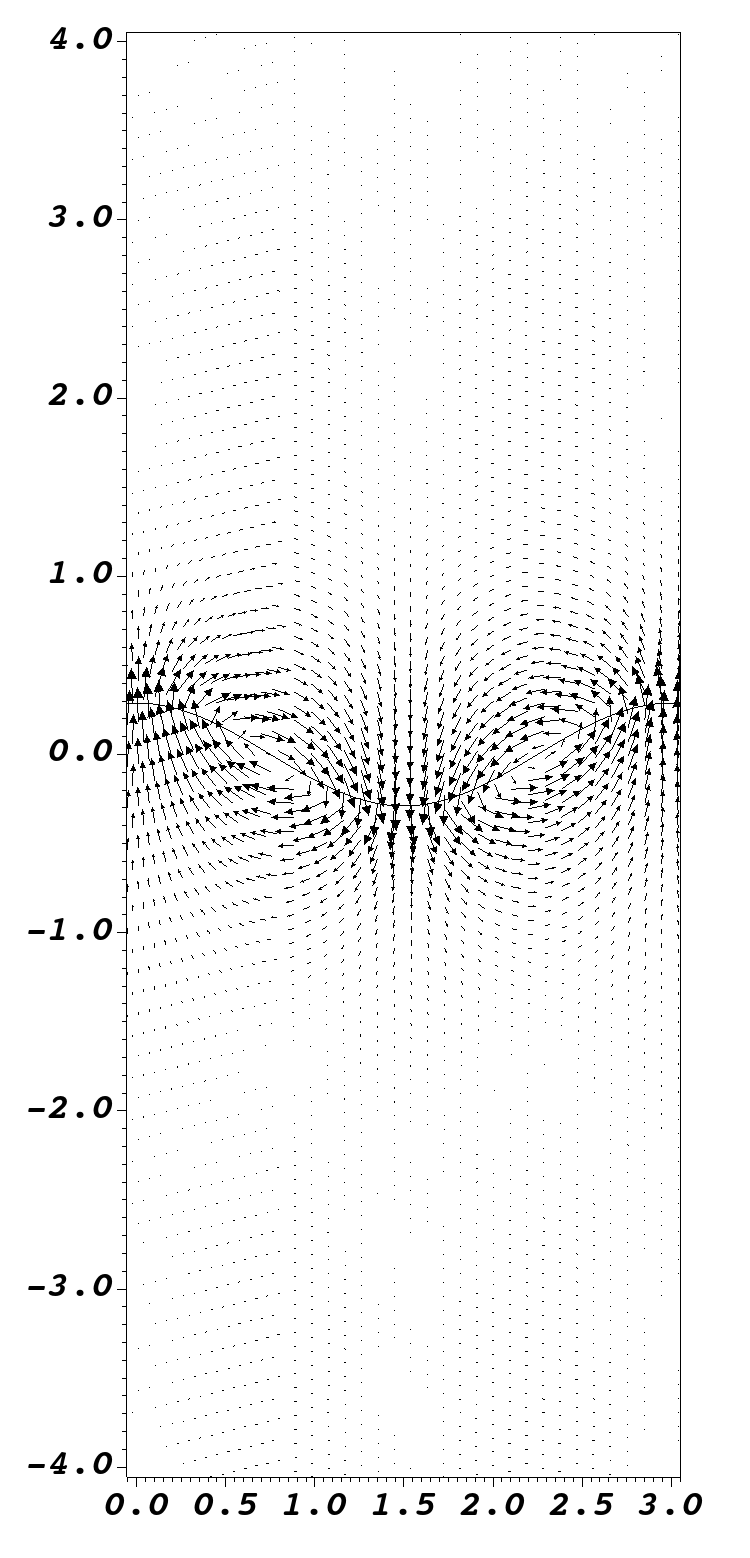}
\includegraphics[scale=.42]{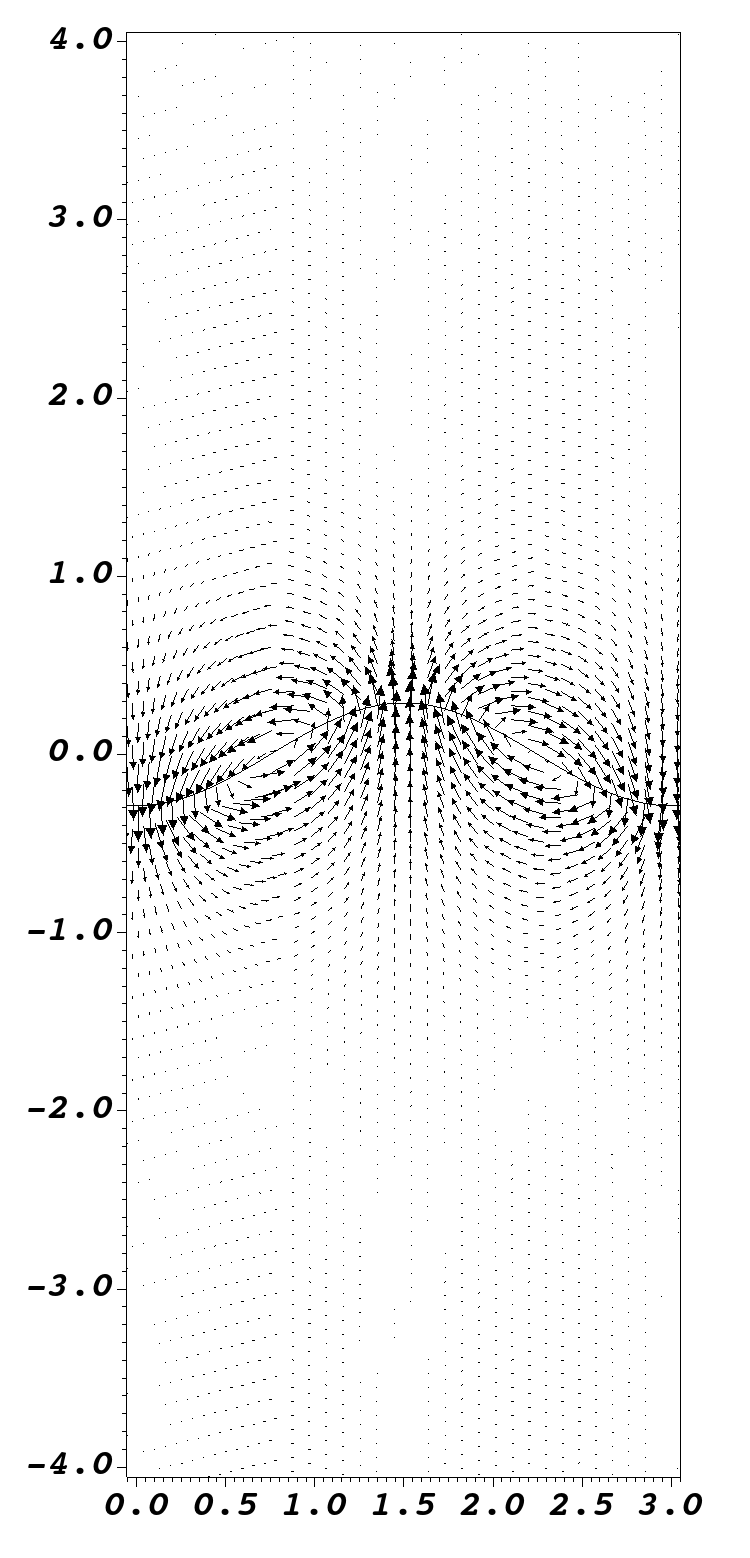}
\caption{Velocity vector field for the spike (left) and bubble (right) on a grid resolution 
$192\times 512$ in the linear regime at early time $t=16.7~ms$.}
\label{fig:velField-early}
\end{figure}

\begin{figure}
\centering
\includegraphics[scale=.42]{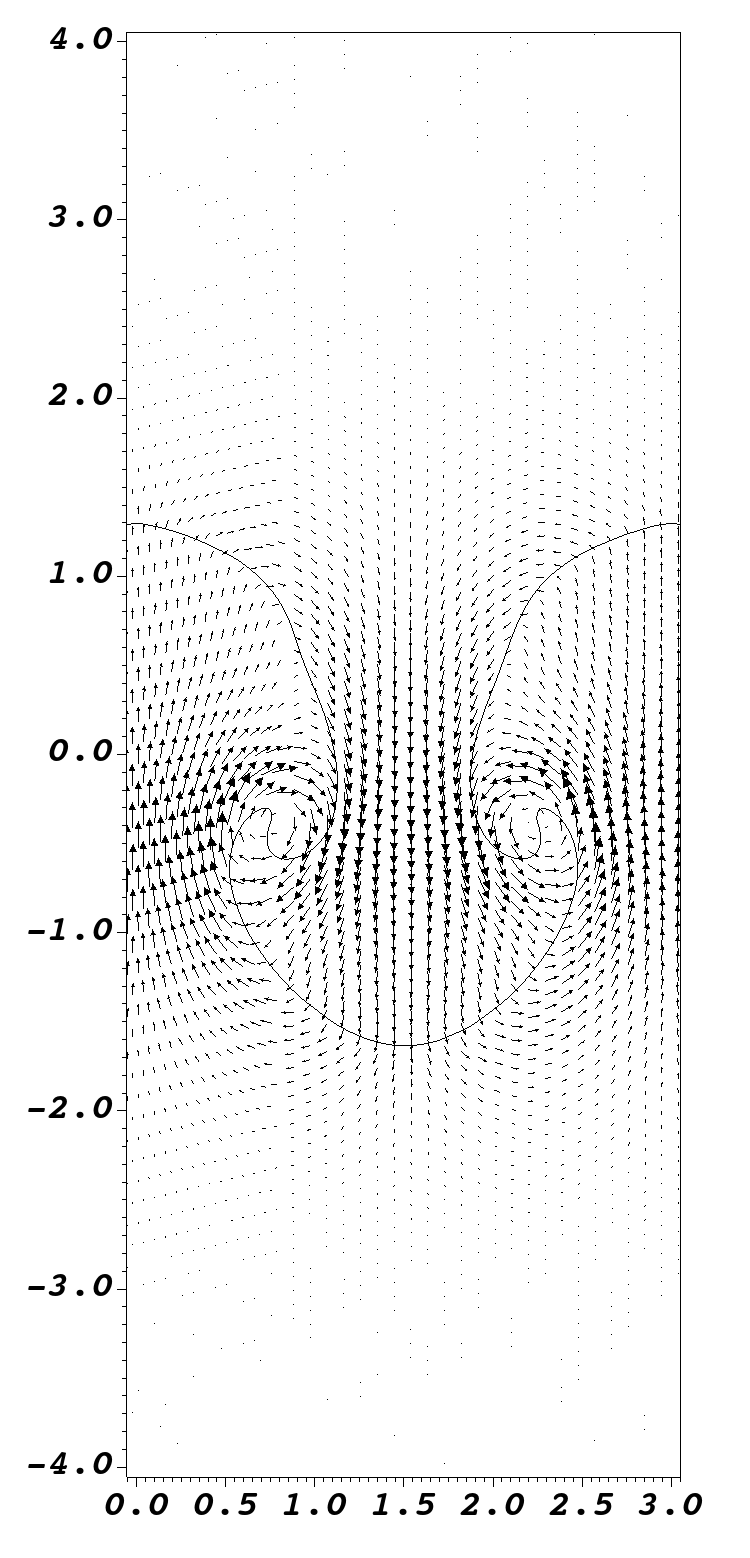}
\includegraphics[scale=.42]{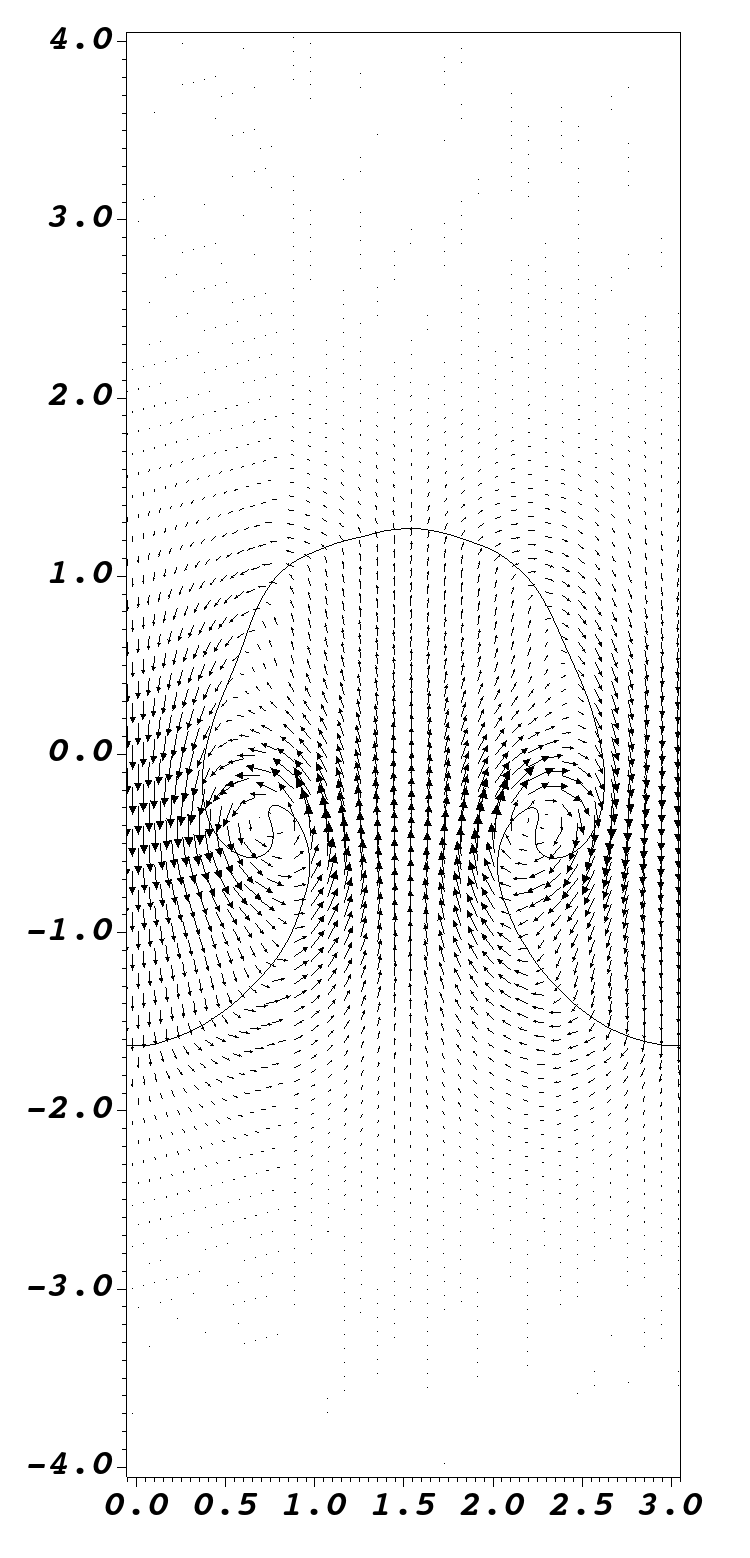}
\caption{Velocity vector field for the spike (left) and bubble (right) on a grid resolution 
$192\times 512$ in the non-linear regime at late time $t=153$~ms.}
\label{fig:velField-late}
\end{figure}

Figs.~\ref{fig:velField-early} and~\ref{fig:velField-late} show the velocity vector field for the spike (left) and bubble (right) in the linear and nonlinear regimes, respectively. 
As expected, rotational motion near the interface is generated from a misalignment of the pressure and density gradients. We observe rotational motion at the interface and no motion away from the interface, which is consistent with the theory presented in~\cite{ChaJaiHwa23}. 

\section{Concluding Remarks}
\indent
The numerical methods used in this study are front-tracking (FT), 
ghost-fluid method (GFM) and high-order weighted essentially non-oscillatory (WENO). 
FT enforces the exact jump conditions at the tracked fluid interface by solving Riemann solvers.
GFM avoids the unphysical oscillations at the fluid interfaces by separating the domain for each fluid and defining each fluid domain with a ghost-fluid region.
WENO ensures that maximum order of accuracy is achieved in the regions away from the interface. 
These methods are implemented in \textsc{FronTier} code to solve 
compressible Euler equations. 
For the validation and verification of \textsc{FronTier}, 
Renoult~{\it et al.}~\cite{RenRosCar15} experiments serve as benchmark tests. 
The interface dynamics, amplitude growth, and asymmetry on the bubble/spike structures of our simulations show good agreement with the experiments. The velocity vector fields for the bubble and spike in early and late times 
are consistent with the theory, which shows rotational motion at the interface and no motion away from the interface.

\begin{acknowledgement}
Tulin~Kaman acknowledges the support of the Verena Meyer Visiting Professorship, 
funded by the Faculty of Business, Economics and Informatics, 
in the Department of Informatics at the University of Zurich and
the Lawrence Jesser Toll Jr. endowed chair in the Department of Mathematical Sciences at the University of Arkansas.
This material is based on work supported by the National Science Foundation under Grant Number OAC-2346752.
This research is supported by the Arkansas High Performance Computing Center (AHPCC) which is funded through multiple National Science Foundation grants and the Arkansas Economic Development Commission.
This material is based upon work supported by the National Science Foundation under Grant No. DMS-1929284 while the authors were in residence at the Institute for Computational and Experimental Research in Mathematics in Providence, Rhode Island, during the ``Modeling and Simulations in Fluids'' hot topics workshop. This workshop honors Dr.~James~Glimm for his outstanding contributions to computational fluid dynamics.
\end{acknowledgement}

\ethics{Competing Interests}{
The authors have no conflicts of interest to declare that are relevant to the content of this chapter.}

\bibliographystyle{plain}
\bibliography{refs}

@String{IJNAM =  "Int. J. Numer. Anal.Mod."}

@String{JCP =    "J. Comput. Phys."}

@article{Aba10,
author = {Abarzhi, Snezhana I. },
title = {Review of theoretical modelling approaches of Rayleigh–Taylor instabilities and turbulent mixing},
journal = {Philosophical Transactions of the Royal Society A: Mathematical, Physical and Engineering Sciences},
volume = {368},
number = {1916},
pages = {1809-1828},
year = {2010},
}

@article{BalShu00,
author = {D. S. Balsara and C. W. Shu},
title = {Monotonicity Preserving Weighted Essentially Non-oscillatory Schemes with Increasingly High Order of Accuracy},
journal = JCP,
volume = {160},
number = {2},
pages = {405-452},
year = {2000},
}

@article{Ban20,
    author = {Banerjee, Arindam},
    title = {Rayleigh-Taylor Instability: A Status Review of Experimental Designs and Measurement Diagnostics},
    journal = {Journal of Fluids Engineering},
    volume = {142},
    number = {12},
    pages = {120801},
    year = {2020},
}

@Article{BoLiuGli10,
  author =       "W. Bo and X. Liu and J. Glimm and X. Li",
  title =        "A robust front tracking method: Verification and
                 application to simulation of the primary breakup of a
                 liquid jet",
  journal =      "SIAM J. Sci. Comput.",
  volume =       "33",
  pages =        "1505--1524",
  year =         "2011",
}

@article{ChaJaiHwa23,
    author = {Chan, Wai Hong Ronald and Jain, Suhas S. and Hwang, Hanul and Naveh, Annie and Abarzhi, Snezhana I.},
    title = {Theory and simulations of linear and nonlinear two-dimensional Rayleigh–Taylor dynamics with variable acceleration},
    journal = {Physics of Fluids},
    volume = {35},
    number = {4},
    pages = {042109},
    year = {2023},
}

@article{ColJac02, 
author={Collins, B. D. and Jacobs, J. W.}, 
title={{PLIF flow visualization and measurements of the Richtmyer–Meshkov instability of an air/SF6 interface}}, 
volume={464}, 
journal={Journal of Fluid Mechanics}, 
publisher={Cambridge University Press}, 
year={2002}, 
pages={113–136}}

@Article{FedOsh99,
  author =       "R. P. Fedkiw and T. Aslam and B. Merriman and S.
                 Osher",
  title =        "A Non-Oscillatory {E}ulerian Approach to Interfaces in
                 Multimaterial Flows (The Ghost Fluid Method)",
  journal =      JCP,
  volume =       "152",
  pages =        "457--492",
  year =         "1999",
}

@article{GliCheSha20,
author = {J. Glimm and B. Cheng and D. H. Sharp and T. Kaman},
title = {A crisis for the verification and validation of turbulence simulations},
journal = {Physica D: Nonlinear Phenomena},
volume = {404},
pages = {132346},
year = {2020},
}

@Article{GliGroLi99a,
  author =       "J. Glimm and J. W. Grove and X.-L. Li and D. C. Tan",
  title =        "Robust Computational Algorithms for Dynamic Interface
                 Tracking in Three Dimensions",
  journal =      "SIAM J. Sci. Comput.",
  volume =       "21",
  pages =        "2240--2256",
  year =         "2000",
}

@Article{GliShaKam11,
  author =       "J. Glimm and D. H. Sharp and T. Kaman and H. Lim",
  title =        "New Directions for {R}ayleigh-{T}aylor Mixing",
  journal =      "Phil. Trans. R. Soc. A",
  volume =	 "371",
  pages =        "20120183",
  year =         "2013",
  note =         "{L}os Alamos National Laboratory Preprint
                 LA-UR 11-00423 and Stony Brook University Preprint
                 SUNYSB-AMS-11-01",
  doi =		 "http://dx.doi.org/10.1098/rsta.2012.0183",
  MRnumber =     "3123296",
}

@Article{KamGliSha10,
  author =       "T. Kaman and J. Glimm and D. H. Sharp",
  title =        "Initial Conditions for Turbulent Mixing Simulations",
  journal =      "Condensed Matter Physics",
  volume =       "13",
  pages =        "43401",
  year =         "2010",
}

@Article{KamMelRao11,
  author =       "T. Kaman and J. Melvin and P. Rao and R. Kaufman and
                 H. Lim and Y. Yu and J. Glimm and D. H. Sharp",
  title =        "Recent Progress in Turbulent Mixing",
  journal =      "Physica Scripta",
  year =         "2013",
  pages =        "014051",
}

@article{KamHol22,
author = {Kaman, T. and Holley, R.},
year = {2022},
volume={19},
number={6},
pages = {822-838},
title = {Validation and Verification of Turbulence Mixing due to {R}ichtmyer-{M}eshkov Instability of an air/{SF6} interface},
journal={IJNAM}
}

@article{HolKam25,
author = {Holley, R. and Kaman, T.},
year = {2025},
volume={},
number={},
pages = {},
title = {A front-tracking/ghost-fluid method for the numerical simulations of {R}ichtmyer--{M}eshkov {I}nstability},
journal={Physica D: Nonlinear Phenomena}
}

@Article{LiuKhoWan05,
  author =       "T. G. Liu and B. C. Khoo and C. W. Wang",
  title =        "The ghost fluid method for compressible gas-water simulation",
  journal =      "J. Comput. Phys.",
  volume =       "204",
  number =       "1",
  year =         "2005",
  ISSN =         "0021-9991",
  pages =        "193--221",
  publisher =    "Academic Press Professional, Inc.",
  address =      "San Diego, CA, USA",
}

@PhdThesis{Mue08,
  author =       "Nicholas J. Mueschke",
  title =        "Experimental and numerical study of molecular mixing
                 dynamics in {R}ayleigh-{T}aylor unstable flows",
  school =       "Texas A and M University",
  year =         "2008",
}

@Article{RamAnd04,
  author =       "P. Ramaprabhu and M. Andrews",
  title =        "Experimental Investigation of {Rayleigh}-{Taylor}
                 Mixing at Small Atwood Numbers",
  journal =      "J. Fluid Mech.",
  volume =       "502",
  pages =        "233--271",
  year =         "2004",
}

@Article{Rea84,
  author =       "K. I. Read",
  title =        "Experimental Investigation of Turbulent Mixing by
                 {Rayleigh}-{Taylor} Instability",
  journal =      "Physica D",
  volume =       "12",
  pages =        "45--58",
  year =         "1984",
}

@article{RenRosCar15,
  title = {Nodal Analysis of Nonlinear Behavior of the Instability at a Fluid Interface},
  author = {Renoult, M.-C. and Rosenblatt, C. and Carles, P.},
  journal = {Phys. Rev. Lett.},
  volume = {114},
  issue = {11},
  pages = {114503},
  year = {2015},
 }

@article{Shu20, 
title={Essentially non-oscillatory and weighted essentially non-oscillatory schemes}, 
volume={29}, 
journal={Acta Numerica}, 
publisher={Cambridge University Press}, 
author={Shu, C.-W.}, 
year={2020}, 
pages={701–762}
}

@TechReport{SmeYou87,
  author =       "V. S. Smeeton and D. L. Youngs",
  type =         "AWE Report Number",
  number =       "0 35/87",
  title =        "Experimental investigation of turbulent mixing by
                 {Rayleigh}-{Taylor} instability (Part 3)",
  year =         "1987",
}

@Article{TerTry09,
  author =       "H. Terashima and G. Tryggvason",
  title =        "A front-tracking/ghost-fluid method for fluid
                 interface in compressible flows",
  journal =      JCP,
  volume =       "228",
  pages =        "4012--4037",
  year =         "2009",
}

@Article{ZhaKamShe18,
  author =	"H. Zhang and T. Kaman and D. She and B. Cheng and
  		J. Glimm and D. H. Sharp",
  title =	"{V}\&{V} for turbulent mixing in the intermediate
  		asymptotic regime",
  year = 	"2018",
  volume =	"14",
  pages =	"193-222",
  journal =	"Pure and Applied Mathematics Quarterly",
  note =	"Los Alamos National Laboratory preprint LA-UR-18-22134 ",
}

@Article{Zho17a,
  Author =	 "Y. Zhou",
  title =	 "{R}ayleigh-{T}aylor and {R}ichtmyer-{M}eshkov instability
  		  induced flow, turbulence, and mixing {I}",
  journal =	 "Physics Reports",
  volume =	 "720--722",
  pages =	 "1--136",
  year =	 "2017",
}

@Article{Zho17b,
  Author =	 "Y. Zhou",
  title =	 "{R}ayleigh-{T}aylor and {R}ichtmyer-{M}eshkov instability
  		  induced flow, turbulence, and mixing {II}",
  journal =	 "Physics Reports",
  volume =	 "723--725",
  pages =	 "1--160",
  year =	 "2017",
}

@Book{Zho24,
  author =       "Y. Zhou",
  title =        "Hydrodynamic Instabilities and Turbulence",
  subtitle =     "Rayleigh–Taylor, Richtmyer–Meshkov, and Kelvin–Helmholtz Mixing",
  publisher =    "Cambridge University Press",
  address =      "Cambridge, England",
  year =         "1993",
}
\end{document}